\documentclass[doublecol,figures]{epl2} 
\usepackage[apple mac]{inputenc}
\usepackage{graphicx}
\usepackage{epstopdf}
\usepackage{picins}
\usepackage{epsf}
\usepackage{subfigure}
\usepackage{amsmath}
\usepackage{amssymb}
\usepackage{amsfonts}

\title{ Quantum Hall-like effect for cold atoms in non-Abelian gauge potentials}

\author{ N. Goldman \and P. Gaspard}
\institute{ Center for Nonlinear Phenomena and Complex Systems -
Universit\'e Libre de Bruxelles (U.L.B.), Code Postal 231, Campus Plaine, B-1050
Brussels, Belgium}

\pacs{03.65.Vf}{Phases: geometric; dynamic or topological}
\pacs{03.75.Ss}{Degenerate Fermi gases}
\pacs{05.60.Gg}{Quantum transport}

\abstract{
We study the transport of cold fermionic atoms trapped in optical lattices in the presence of artificial Abelian or non-Abelian gauge potentials. Such external potentials can be created in optical lattices in which atom tunneling is laser assisted and described by commutative or non-commutative tunneling operators.  We show that the Hall-like transverse conductivity of such systems is quantized by relating the transverse conductivity to topological invariants known as Chern numbers.  
We show that this quantization is robust in non-Abelian potentials. The
different integer values of this conductivity are explicitly computed
for a specific non-Abelian system which leads to a fractal phase
diagram.} 

\begin{document}

\maketitle


During the last decades, remarkable quantum phenomena have been discovered such as 
the Aharonov-Bohm effect \cite{aharonov1959}, the geometric phases \cite{berry1984,Mead1992}, and the integer quantum Hall effect (IQHE) \cite{Klitzing1986,Thouless1982}. These important phenomena manifest themselves 
when gauge potentials are present in non-trivial topological spaces, in which cases the wave functions describing the system may acquire topological or geometrical phases. 
They are elegantly described in terms of differential geometry and topology in which the topological phases are related to the concept of holonomy \cite{Simon1983}.  In the IQHE, Hall's transverse conductivity of two-dimensional electronic systems submitted to high magnetic fields turns out to be quantized according to
$\sigma _H =-C \, (e^2/h)$ where $C$ is a topologically invariant integer called the first Chern number \cite{Simon1983,Avron1983,Nakahara}. Since the proportionality factor is a combination of the electric charge $e$ and Planck's constant $h$, such quantization phenomena play a key role in metrology \cite{Klitzing1986}.

Very recently, great experimental advances have been performed on the control of cold atomic systems 
\cite{Chin2006,Kohl2005 ,Gunter2005,Stoferle2006}.  In such systems, artificial gauge potentials can be 
created in optical lattices in order to reproduce the dynamics of periodically constrained fermionic atoms submitted to the analogue of a magnetic field. Such systems would thus offer the possibility to observe the corresponding fractal energy spectrum known as the Hofstadter butterfly \cite{Hofstadter1976,Zoller2003}.
More recently the realization of non-Abelian gauge potentials has also been envisaged, allowing the observation of a non-Abelian Aharonov-Bohm effect \cite{Osterloh2005}, magnetic monopoles \cite{Ruseckas2005}, and particular metal-insulator transitions \cite{Satija2006}.  

The purpose of this Letter is to report the possibility of an integer quantum Hall-like effect in the transport
of cold fermionic atoms trapped in optical lattices submitted to artificial gauge potentials.  We consider this effect in a family of non-Abelian gauge potentials, which contains Abelian potentials as particular cases.
In spite of the new topological context provided by the non-Abelian gauge structure, we show that
an integer quantum Hall-like effect is possible and that the transverse conductivity is indeed quantized in terms of Chern numbers in systems of fermionic cold atoms in optical lattices with artificial Abelian 
or non-Abelian gauge potentials.
Since the current induced in our system is electrically neutral, this quantized conductivity has the distinctive units of the inverse of Planck's constant, which could be important for metrological purposes. 
In this Letter, we first define the system Hamiltonian characterized by its non-Abelian gauge structure 
and evaluate the transverse conductivity with Kubo's formula which gives the linear response of
the system to an external static perturbation. We then  express the
Hall-like conductivity in terms of Chern numbers in order to prove the quantization of this physical quantity. We compute these topological invariants for a specific system which leads to a fractal phase diagram. 
We then discuss possible experimental setups for the validation of our theoretical considerations.

We are interested in systems described by the following Hamiltonian
\begin{equation}
\mathcal{H}= \sum_{m,n} t_a \,  \, \{ c^{\dagger}_{m,n} \vert U_x
\vert c_{m+1,n} \} + t_b  \,  \, \{ c^{\dagger}_{m,n} \vert U_y \vert
c_{m,n+1} \} + {\rm h.c.}
\label{ham}
\end{equation}
which determines the dynamics of fermions in a two-dimensional
lattice within a tight-binding approximation as studied
experimentally in cold atom systems trapped in optical lattices  \cite{Chin2006,Kohl2005,Gunter2005,Stoferle2006}. For
$U_{x,y}$ belonging to the Abelian group of unitary complex numbers $U(1)$, this Hamiltonian reproduces the evolution of an
electronic system constrained by a periodic potential and submitted
to a magnetic field \cite{Zoller2003}. In the latter case, the single-particle wave function gets a phase factor $U_x$ ($U_y$) when tunneling is performed along the $x$ ($y$) direction and their product around a unit cell gives a total phase proportional to the penetrating magnetic flux. In this Letter, we study the case in which the
system has a $U(2)$ gauge structure, which implies that the operators
$U_x$ and $U_y$ are $2 \times 2$ unitary matrices and the single-particle wave function has two components: when tunneling occurs along the $x$ ($y$) direction, the matrix $U_x$ ($U_y$) acts as a tunneling operator on the two-component wave function and the system is non-Abelian for $[U_x,U_y] \ne 0$ \cite{Osterloh2005,Satija2006}. The operators $U_{x,y}$
are related to the gauge potential $\boldsymbol{A}$ present in the system via the usual relation $U_{x,y}=e^{i A_{x,y}}$.
In order to indicate multicomponent  wave functions and operators, we use the following notations
introduced in Ref. \cite{Mead1992}: A \emph{row} of $n$ orthogonal kets
 is denoted by $\{ \vert \psi \rangle \vert=(\vert \psi
\rangle_1, ... , \vert \psi \rangle _n)$ and a \emph{column}
of $n$ kets by $\vert \vert \psi \rangle
\}$.The coefficients
$t_a$ and $t_b$ describe transfers in the $x=m l_x$ and $y=n l_y$
directions, where $(l_x, l_y)$ are the unit cell's lengths, and $\vert c^{\dagger}_{m,n} \}$ (resp. $\vert c_{m,n}
\}$) are the $2$-component fermionic creation (resp. annihilation)
field operators on site $(m,n)$. After a change of basis, the latter's components are transformed into
$c_{m,n,j}=\sum_{\lambda}  \psi_{\lambda j}(m,n) \, c_{\lambda} $,
where $c_{\lambda}$ is the operator of annihilation of a fermion in
the $j^{\textrm{th}}$ component $\psi_{\lambda j}(m,n)$ of the
single-particle wave function $\{\psi_{\lambda}(\boldsymbol{r})
\vert=(\psi_{ \lambda 1}(\boldsymbol{r}),\psi_{ \lambda
2}(\boldsymbol{r}))$ with $\boldsymbol{r}=(m,n)$ and $j=1,2$. 

The current density of the system $\boldsymbol{j}_{m,n}$ associated with the
Hamiltonian operator \eqref{ham} is written in terms of the fermionic operators
and its components are given by
\begin{align}j_{m,n,x} =&\frac{i t_a}{\hbar}
\sum_{\lambda\lambda'jj'}\psi ^*_{ \lambda j}(m-1,n)  (U_x)_{_{jj'}}
\psi_{\lambda ' j'} (m,n)  c_{\lambda} ^{\dagger}  c_{\lambda ' } \notag \\ 
& +{\rm h.c.} \notag
\end{align} 
and a similar expression for $j_{m,n,y}$.

It has been recently suggested that non-Abelian gauge potentials can be created in optical lattices \cite{Osterloh2005,Satija2006}. In such setups, atoms with doubly degenerate Zeeman sublevels hop from a site to another with the assistance of additional lasers and their tunnelings are indeed described by unitary  non-commutative operators \cite{Osterloh2005}.  The corresponding gauge potential is given by
 \begin{equation}
\boldsymbol{A}=  \Biggl [\begin{pmatrix} -\frac{\pi}{2}
&\frac{\pi}{2}e^{i \phi} \\
\frac{\pi}{2}e^{-i \phi} &-\frac{\pi}{2}
\end{pmatrix},\begin{pmatrix} 2 \pi m \alpha _1 &0 \\
0 &2 \pi m \alpha _2 \end{pmatrix},0 \Biggr]
\label{potential} \end{equation} and induces an effective
``magnetic" field characterized by three parameters $\alpha _1$, $\alpha _2$ and $\phi$,  which can be controlled with external lasers \cite{Osterloh2005}.   We notice that this gauge potential is Abelian under the condition that $\alpha_1-\alpha_2$ is an integer and non-Abelian otherwise.

We evaluate the linear response of the system to
an external static force applied along the $y$ direction
$\mathcal{H}_{\rm ext}= -\Delta _y \sum_{m,n} n \, \{
c^{\dagger}_{m,n} \vert c_{m,n} \}$. A generalized Lorentz force
produced by the effective ``magnetic" field induces a transverse
current which satisfies $\langle j_x \rangle = \sigma _{x y} \,
\Delta _y $. Kubo's formula expresses the transverse conductivity in
terms of the total current $\boldsymbol{J} \equiv
\sum_{m,n}\boldsymbol{j}_{m,n}$
\begin{equation}
\sigma _{x y}=\frac{i\hbar}{V}\sum_b \frac{\langle b \vert J_{x}
\vert 0 \rangle \langle 0 \vert J_{y} \vert b \rangle}{(E_b
-E_0)^2}-\frac{\langle 0 \vert J_{x} \vert b \rangle \langle b \vert
J_{y} \vert 0 \rangle}{(E_b -E_0)^2}
\end{equation}where $\vert b\rangle$ denotes the eigenstates of the
Hamiltonian (\ref{ham}) of eigenvalue $E_b$ and $V$ is the system
volume or area.

The system presents symmetries under translations defined by the
operators $ \vert \bigl ( T^q_x \, \psi \bigr ) (m,n) \} = \vert
\psi (m+q,n) \}$ and
$ \vert \bigl (  T_y \, \psi \bigr ) (m,n) \} =  \vert \psi (m,n+1)
\} $. The operator $T_y$ commutes with the single-particle
Hamiltonian $H$, since the potential (\ref{potential}) only depends
on the $x$ coordinate. On the other hand, the commutation relations
$[T_x^q,H]=[T_x^q,T_y]=0$ are satisfied under the conditions
$\alpha_j=p_j/q$ with $j=1,2$ and $p_1,p_2,q$ some integer numbers.
The translational symmetries allow us to write $\vert \psi (m,n) \} =
e^{i \boldsymbol{k} \cdot \boldsymbol{r}} \vert u (m)\}
\label{bloch}$ with $\vert u (m)\} $ $q$-periodic and the first
Brillouin zone is a $2$-torus $\mathbb{T}^2$ defined by $k_x \in [0 ,
\frac{2 \pi}{q}]$ and $k_y \in [-\pi,\pi]$. The latter change can be
viewed as a gauge transformation which modifies the Hamiltonian by
substituting $A_{\mu} \rightarrow A_{\mu}' = A_{\mu} + k_{\mu}$ so
that the single-particle Hamiltonian takes the following form
\begin{align}
&H(\boldsymbol{k}) \vert u (m) \} = \notag \\
& t_a e^{i(A_x +  k_x)} \vert u(m+1) \} + t_a e^{- i (A_x +  k_x)}
\vert u(m-1) \} \notag \\
&+ t_b e^{i(A_y + k_y)} \vert u(m) \}  +  t_b e^{-i (A_y + k_y)}
\vert u(m) \} \label{ham1}
\end{align}
This equation can be differentiated with respect to $k_{\mu}$ and a
straightforward calculation yields
\begin{align}
\sum_{b} \frac{\langle b \vert J_{x} \vert 0 \rangle \langle 0 \vert
J_{y} \vert b \rangle}{(E_b -E_0)^2} &= \frac{1}{\hbar ^2}
\sum_{\epsilon _{\lambda} < \epsilon _{\rm F}} \sum_{\epsilon
_{\lambda '} >\epsilon _{\rm F}}  \{ \langle u_{\lambda '}  \vert
\partial_{k_{x}} H \vert u_{\lambda}  \rangle \} \notag \\
& \times \{ \langle u_{\lambda}  \vert \partial_{k_{y}} H \vert
u_{\lambda '} \rangle \}/(\epsilon_{\lambda}-\epsilon_{\lambda'})^2
\label{iden}\end{align}
where $H\vert u_{\lambda}\}=\epsilon_{\lambda}\vert u_{\lambda}\}$
are the single-particle eigenstates and the scalar product is defined
by $\{ \langle u_{\lambda}  \vert u_{\lambda '}  \rangle \}
=\sum_{m=1}^q \sum_{j=1,2} u_{\lambda j}^* (m) \,  u_{\lambda ' j}
(m)$. The Fermi energy $\epsilon _{\rm F}$ is supposed to be situated
inside a gap of the spectrum. We find that $\sigma_{xy}=-\sigma_{yx}$ and the identity (\ref{iden}) allows us to
write the conductivity as
\begin{align}
\sigma_{x y}&= \frac {1}{(2 \pi )^2 i \hbar}  \sum_{\epsilon _{\lambda}
< \epsilon _{\rm F}}  \int_{\mathbb{T}^2} \sum_j \Big[ \langle
\partial _{k_{x}} u_{\lambda j} \vert \partial _{k_{y}} u_{ \lambda
j} \rangle  \notag \\ \notag &\qquad\qquad\qquad\qquad -\langle
\partial _{k_{y}} u_{\lambda j} \vert \partial _{k_{x}} u_{ \lambda
j} \rangle  \Big] d\boldsymbol{k} \\
\label{hal}
\end{align}
This result is important because it relates the transverse conductivity to integer topological invariants called Chern numbers. The latter are defined on a \emph{fibre bundle} which is a topological space that locally resemble the direct product of the parameter space $\mathbb{T}^2$ with the non-Abelian gauge group $U(2)$ and that we note $P(\mathbb{T}^2, U(2))$. This product space may be globally non-trivial and the structure \emph{twists} when the Chern number is non-zero. In the Abelian framework, the interpretation of the IQHE in terms of topological arguments \cite{Simon1983,Avron1983} showed that Berry's curvature $\mathcal{F}= \bigl[ \langle
\partial _{k_{x}} \psi (\boldsymbol{k} ) \vert \partial _{k_{y}}  \psi (\boldsymbol{k} ) \rangle - \langle
\partial _{k_{y}} \psi (\boldsymbol{k} ) \vert \partial _{k_{x}}  \psi (\boldsymbol{k} )  \rangle \bigr] dk_{x} dk_{y}$, with $\vert \psi \rangle$ an eigenstate depending on the wave vector $\boldsymbol{k}$, defines a curvature on the fibre bundle $P(\mathbb{T}^2, U(1))$ and that its integral over the parameter space $\mathbb{T}^2$ corresponds to the first Chern number $C= \frac{i}{2 \pi} \int_{\mathbb{T}^2} \mathcal{F} $, which is an integer. These concepts can be generalized in the actual non-Abelian framework \cite{Hatsugai2004,Hatsugai2006}  for which Berry's curvature $\mathcal{F}$ is a $2 \times 2$ matrix  and the Chern number is then defined by  
\begin{equation}
C= \frac{i}{2 \pi} \int_{\mathbb{T}^2} {\rm tr}  \bigl ( \vert \langle
\partial _{k_{x}} u \vert \} \{ \vert \partial _{k_{y}} u \rangle \vert -\vert \langle
\partial _{k_{y}} u \vert \} \{ \vert \partial _{k_{x}} u \rangle \vert \bigr ) dk_{x} dk_{y}
\end{equation}
where ${\rm tr}\mathcal{F}$ denotes the trace of the matrix $\mathcal{F}$. One eventually finds that the Hall-like conductivity \eqref{hal} is given by a sum of Chern numbers
\begin{align}
\sigma_{x y}&= -\frac {1}{h}  \sum_{\epsilon _{\lambda}
< \epsilon _{\rm F}}  C(\epsilon_{\lambda}) 
\end{align}
where the integer numbers $C(\epsilon_{\lambda})$ are associated with each energy band. It follows that the transverse Hall-like conductivity of the system evolves by steps corresponding to integer multiples of the inverse of Planck's constant and that it remains constant under small perturbations. Besides, a measure of the transverse conductivity in this non-Abelian system provides a direct evaluation of these topological invariants.

The Chern numbers were explicitly obtained in
the usual Abelian framework \cite{Kohmoto1989}, and we here show that
they can be computed explicitly in non-Abelian systems as well.
Using Stokes' theorem one gets the following result
\begin{equation}
\sigma_{x y} =  \sum_{\epsilon _{\lambda} < \epsilon _{\rm F}} \,
\frac{1}{2 \pi h} \oint_{\partial\mathbb{T}^2} \left(\frac{\partial
\theta_{\lambda}}{\partial k_{x}} dk_x+\frac{\partial
\theta_{\lambda}}{\partial k_{y}} dk_y\right)
\label{stokes}\end{equation}
where $ \theta_{\lambda}$ is the phase accumulated by $u_{ \lambda}$
around the torus and evaluated on its border $\partial\mathbb{T}^2$.
These  phases can be evaluated in a way similar as in the Abelian
case \cite{Kohmoto1989}. Let us write the Schr\"odinger equation
associated with the non-Abelian Hamiltonian \eqref{ham1} in the form
\begin{align}
&t_a \begin{pmatrix}
e^{i \phi +i k_x }u_2(m+1) +e^{i \phi - i k_x}u_2(m-1) \\ e^{-i \phi
+ i k_x }u_1(m+1)+e^{-i \phi - i k_x}u_1(m-1) \end{pmatrix} \notag \\
  &+t_b \begin{pmatrix}
2 \cos(2 \pi \alpha _1 m + k_y) u_1(m)\\ 2 \cos(2 \pi \alpha _2 m + k_y) u_2(m)
  \end{pmatrix} = \epsilon \begin{pmatrix} u_1(m) \\
u_2(m)\end{pmatrix} \label{har} \end{align}
When $t_a \rightarrow 0$,  this equation gives a $2 q$-degenerate single band and the
dispersion law for  $u_j(m)$ is
\begin{equation}
\epsilon_j (m, k_y)= 2 t_b \cos(2 \pi p _j m /q +k_y)=:2 t_b \cos(k_y^m)
\label{dis}\end{equation}

First, we consider the case where $\alpha _1= p_1/q$ and $\alpha _2=
(p_1+n q)/q$ with $n$ integer. This case allows the explicit
computation of Chern numbers for Fermi energies situated inside the
various gaps of the spectrum in the approximation of weak coupling
with $t_a$ small and $t_b=1$. Since the Chern numbers are topological
invariants, their weak-coupling values extend to other couplings as
long as the band gaps deform continuously. As $t_a$ increases from $0$
to $1$, different regimes are observed.  For $t_a > 0$ and small,
$q-1$ gaps open at the {\it crossing points} $k_y= \pm \frac{l
\pi}{q}$ with $l$ integer.  Thereafter,  the dispersion branches
separate and no crossing remains.  When
$t_a=1$, the spectra $\epsilon=\epsilon(\alpha_1=\frac{p_1}{q};
\alpha _2=\frac{p_1+n q}{q})$ precisely depict a Hofstadter butterfly 
\cite{Hofstadter1976}. Finally when $t_a>1$, $q$ new gaps open so
that the number of gaps in the spectrum becomes $2 q -1$.

In order to obtain the Chern numbers, one can study how the
eigenfunctions change at the different gaps as $k_y$ varies in the
first Brillouin zone \cite{Kohmoto1989}. Near these gaps, the four functions
$u_{1,2}(m)$ and $u_{1,2}(m')$ with their dispersion branch crossing
each other in the limit $t_a =0$ are strongly coupled together.
Anticrossing appear for $t_a \neq 0$ which can be calculated by
perturbation theory giving an effective Schr\"odinger equation for the
vector $\boldsymbol{v}  (m,m',r)$ formed with these four functions.
Considering the $r^{\rm th}$ gap and setting $t_r=m-m'$, this
equation can be written as $M(q,t_r,k_x) \, \boldsymbol{v} (m,m',r)=
\epsilon \, \boldsymbol{v} (m,m',r)$. The latter system has been
solved in order to find the four eigenvectors $ \boldsymbol{v}_i$.
When $k_y$ passes a {\it crossing point}, the eigenvectors undergo
some transformation $\boldsymbol{v}_i \rightarrow
\tilde{\boldsymbol{v}} _i=R(i,t_r,k_x) \boldsymbol{v} _i$. Following
the eigenvector of the $r^{\rm th}$ band through the successive
anticrossings in the first Brillouin zone $k_y \in [-\pi,\pi]$, one
finds that the eigenvector goes under the total transformation $\vert
u_{\lambda}\} \rightarrow e^{i q (t_r - t_{r-1}) k_x} \vert
u_{\lambda}\}$. We can now evaluate the $r^{\rm th}$ band's
contribution to the transverse transport coefficient $\sigma_{x y} $
by considering the formula \eqref{stokes} which gives $\sigma_{x y}
(r^{\rm th} \, \textrm{band}) =  (t_{r} - t_{r-1})/h$. 

The integers $t_r$ have still to be evaluated. From the definition \eqref{dis}, one
has $\epsilon_1 (m, k_y)=\epsilon_2 (m, k_y)$ in the present
situation where $\alpha _2= (p_1+n q) /q $. Two conditions are
necessary for {\it crossing points} that determine the $r^{\rm th}$
gap when $t_a \ne 0$:
$k_y^{m'}=-k_y^m+2 \pi s_r$ with $s_r$ integer, and
$k_y^{m'}=k_y^m-2 \pi p_1 t_r /q$ with $t_r=m-m'$, which lead to the
Diophantine equation $r= p_1 t_r + q s_r$, with $r$ the position of
the gap and $\alpha _1= p_1 /q , \alpha _2= (p_1+n q) /q $. For
$t_a >1$, all the $2 q-1$ gaps open, so that $\sigma_{xy} =
\sum_{j=1}^r  (t_{j} - t_{j-1})/h=    t_r /h $, if the Fermi energy
lies in the $r^{\rm th}$ gap. For $t_a \le1$, only $q-1$ gaps open
and the bands have a doubled multiplicity.  Accordingly, one finds
$\sigma_{xy}=  2 \, t_r / h $ if the Fermi energy lies in the $r^{\rm
th}$ gap.

\begin{figure}
\hspace{-0.5cm}
\onefigure[scale=0.631]{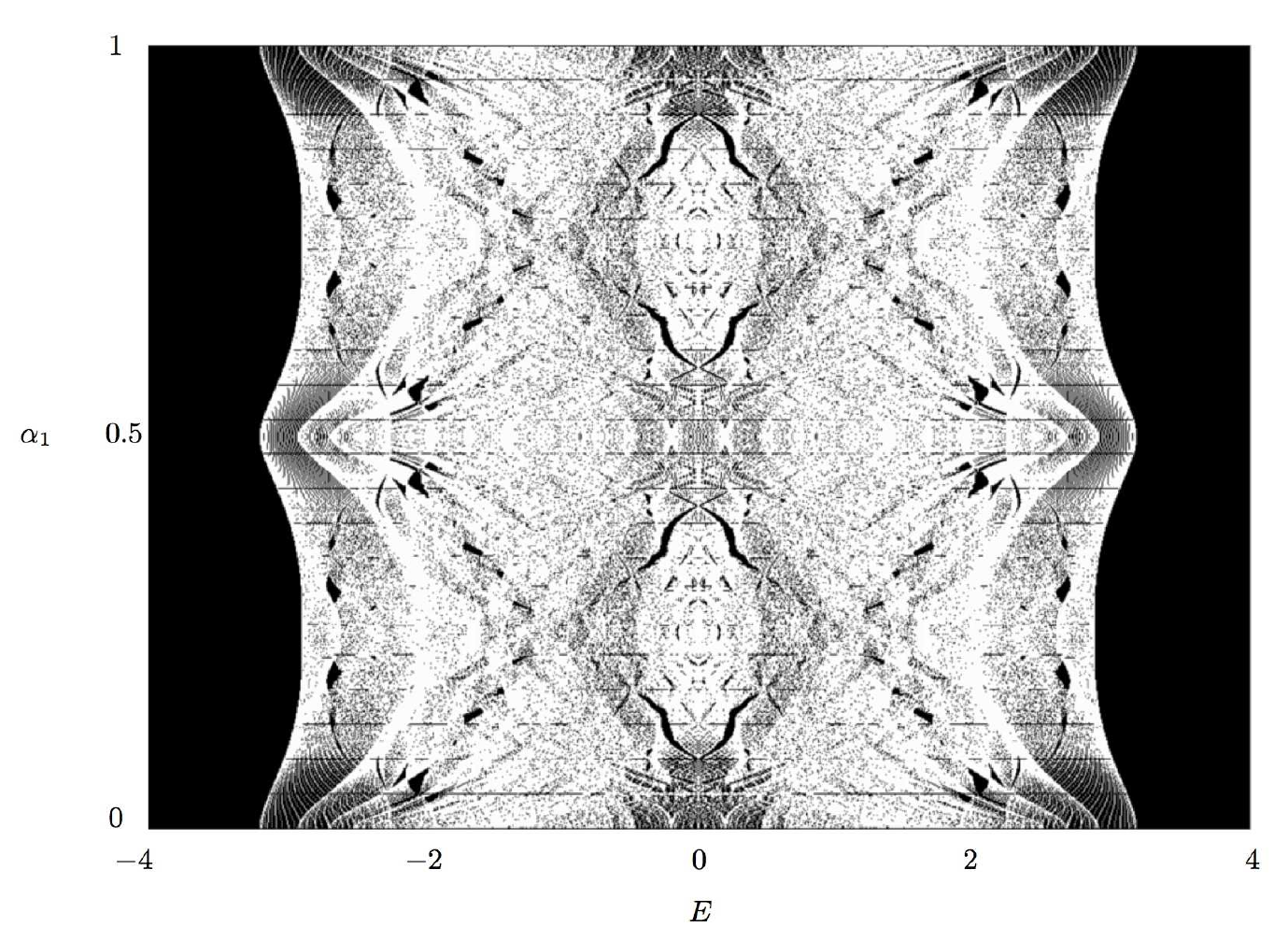}
\caption{\label{d3} Energy spectrum $\epsilon=\epsilon(\alpha _1)$ for $\alpha _2= 31/701$. Gaps are shown in black. A Hofstadter butterfly crosses this structure at $\alpha _1=\alpha _2= 31/701$, where its larger gaps are put forward through thin black lines.}
\end{figure}

\begin{figure*}
\begin{center}
\begin{tabular}{lcr}
\includegraphics[scale=1.2]{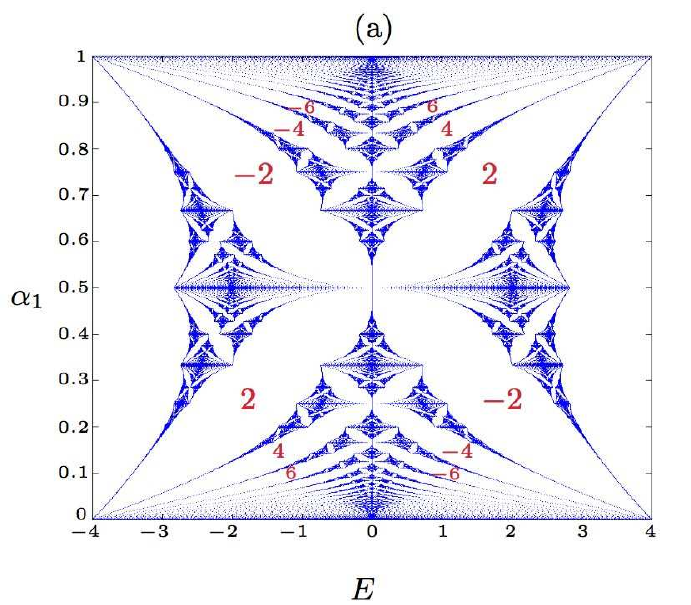} & \includegraphics[scale=1.2]{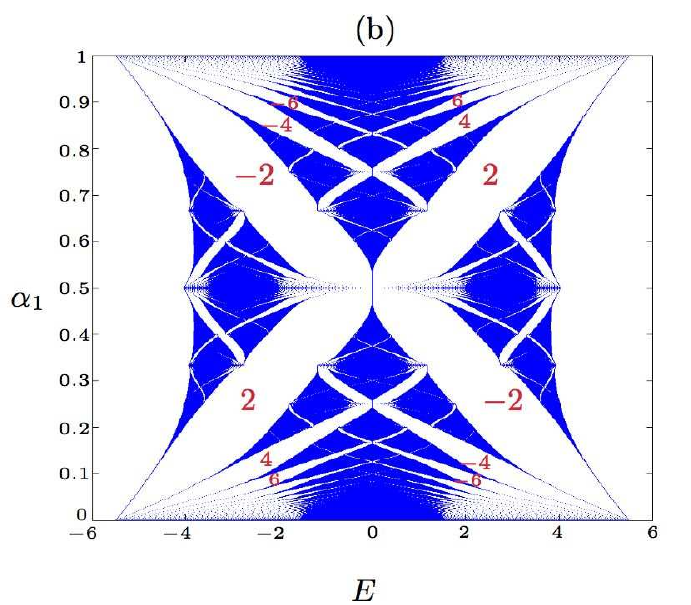} \\
\includegraphics[scale=1.2]{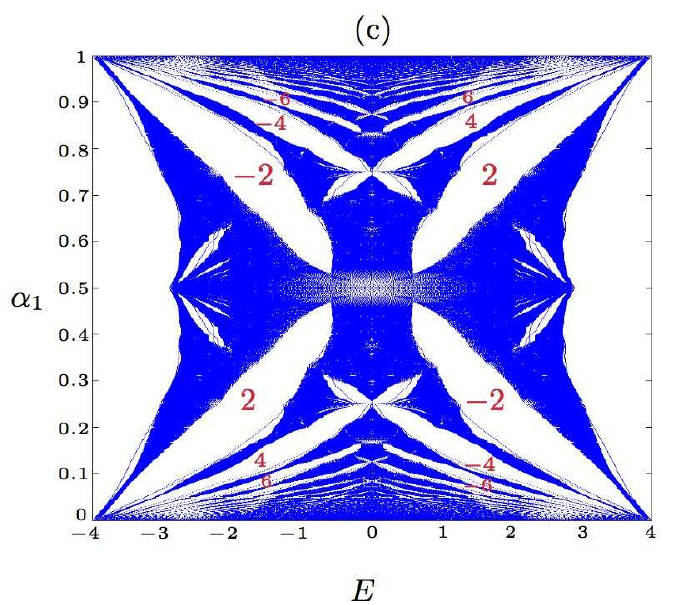} & \includegraphics[scale=1.2]{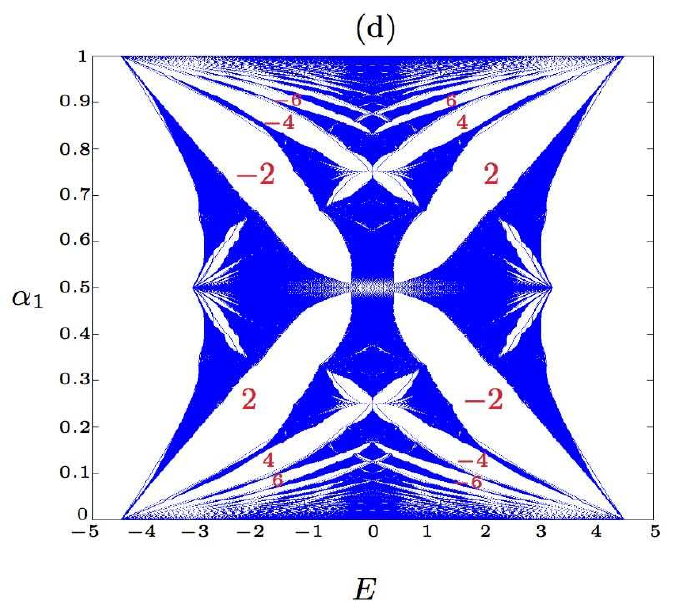} 
\end{tabular}
\caption{\label{c2} Spectra for $\alpha _1= p_1/701$ in the Abelian cases: (a) $\alpha
_2= p_1/701+n$ with $\Lambda = 2 t_b/t_a=2$; (b) $\alpha _2=
p_1/701+n$ with $\Lambda =3.5$; and non-Abelian cases: (c) $\alpha _2= (p_1+0.1)/701+n$ with
$\Lambda =2$; (d)  $\alpha _2= (p_1+0.05)/701+n$ with $\Lambda
=2.5$. The integers in the energy gaps give the corresponding
values of $h \sigma _{y x}$.}
\end{center}
 \end{figure*}

When resolving the non-Abelian Harper equation \eqref{har} ($t_a=t_b=1$) one finds the
energy spectrum $\epsilon=\epsilon(\alpha _1 , \alpha _2)$ which forms, in the $3D$ space ($\epsilon$, $\alpha_1$, $\alpha_2$), a structure called  the
\emph{Hofstadter moth}  \cite{Osterloh2005}. This $3D$ structure is characterized by a complex distribution of small gaps as shown in Fig. \ref{d3} which illustrates a typical cut through the ``moth". If the spectrum
$\epsilon=\epsilon(\alpha_1)$ is depicted in a particular plane  $\alpha_2=(p_1+n
q)/q$ with $n$ integer, one recovers the Hofstadter butterfly \cite{Hofstadter1976, Wannier1978,Petschel1993, Springsguth1997}.  In this spectrum, each gap is characterized by the two integers $s_r$ and $t_r$
given by the aforementioned Diophantine equation \cite{Wannier1978}. 
Consequently, it is possible to draw a phase
diagram similar to Osadchy and Avron's \cite{Osadchy2001} for this
new non-Abelian case. This new phase diagram represents the integer
values of the transverse transport coefficient $\sigma _{x y}$ inside
the infinitely many gaps of the butterfly, which are given by $\sigma
_{x y}=  2 \,  t_r  /h $. In order to check the robustness of this
result, we depict in Fig.   \ref{c2} the energy spectra for different values
of $\alpha_1-\alpha_2$ and anisotropy ratio $t_b/t_a$. The gauge potential
is Abelian in Figs. \ref{c2}(a) and (b), but non-Abelian in Figs. \ref{c2}(c) and (d).
We observe in this latter case that the spectrum thickens in particular around
$\alpha_1=0.5$. Under the effect of the perturbations, 
some gaps disappear as in the Hofstadter model 
\cite{Hasegawa1990}, but the principal gaps where the Hall-like
conductivity is quantized remain.

An important issue is whether this result could be experimentally
probed. A possible system described by the Hamiltonian \eqref{ham}
which includes the gauge potential \eqref{potential} has already been
proposed by Osterloh \emph{et al.} \cite{Osterloh2005} who put
forward the possibility to submit cold fermionic atoms to such
artificial non-Abelian gauge potentials. Each vertex of a $3D$ optical
lattice contains an atom with degenerate Zeeman sublevels in the
hyperfine ground state manifolds $\{\vert g \rangle _j , \vert e
\rangle _j \}$ with $j=1,2$, so that the states we are dealing with are
represented by the rows $\{\vert e \rangle \vert=(\vert e \rangle _1,
\vert e \rangle _2)$ or $\{ \vert g \rangle \vert=(\vert g \rangle
_1, \vert g \rangle _2)$. We note that fermionic atoms with such properties, i.e. $^{40}K$ in states $\vert F=9/2, m_F = 9/2,7/2,... \rangle$ and $\vert F=7/2, m_F = -7/2,-5/2,... \rangle$ as suggested in Ref. \cite{Osterloh2005}, are already studied experimentally in optical lattices \cite{Kohl2005 ,Gunter2005,Stoferle2006}. Lasers are used in order to create the
lattice, as well as the non-Abelian gauge potential through
state-dependent  control of hoppings which take place within every
plane $z=$ constant. In this framework, we consider a possible setup
that exhibits a current in the $y$ direction consisting of an
accelerating optical lattice \cite{Wilkinson1996,Zoller2003}, with the (possibly gravitational)
acceleration $\boldsymbol{a}=(0,a_y)$. Consequently, a perturbative
term $\mathcal{H}_{\rm ext}$ with the intersite energy difference $\Delta _y= M
a_{y} l_y$ where $M$ is the mass of the particles is added to the
Hamiltonian \cite{Zoller2003,Ponomarev2006}. In this case, the
transverse transport coefficient $\sigma _{xy}$ would give the
relation between the external forcing and the transverse atomic
current through the lattice. We notice that the current is electrically neutral so that this conductivity has the distinctive units of the inverse of Planck's constant and relates the energy difference $\Delta _y$ to a number of atoms per unit time in the current. Recently, it has been suggested that
such atomic currents could be experimentally produced and measured in
optical lattices. In particular, Ponomarev \emph{et al.} 
\cite{Ponomarev2006} have investigated theoretically the possibility
of such currents in a system of fermions coupled to bosons. The
latter acts as a cold bath for the fermions, which induces a
relaxation of the Bloch oscillations. If a transverse current is
present in such systems under the conditions described in this work,
we have shown that the transverse Hall-like conductivity should be
quantized. In order to observe experimentally the effects emphasized in this work, it is crucial that the energy gaps represented in Fig. \ref{c2} remain sufficiently clear under thermal effects. The energy resolution required is approximatively $100$ Hz which corresponds to temperatures of the order of $10$ $n$K. These temperatures are realized in experiments involving $^{40}K$ in state $\vert F=9/2, m_F = 9/2 \rangle$ as mentionned in Refs. \cite{Kohl2005 ,Gunter2005,Stoferle2006}. We finally point out that recent studies put forward the possibility to create and measure currents in atomic periodic systems, such as optical lattices, by connecting two reservoirs in order to create a chemical potential gradient through the system which generates a current \cite{Seaman2007}. Such devices, in which the trapping potential present in the optical lattice can be taylored, are envisaged in the new context of \emph{atomtronics} \cite{Seaman2007} and could be considered for the detection of the quantized Hall-like conductivity in optical lattices submitted to Abelian or non-Abelian gauge potentials.

In this Letter, we have shown the
existence of topological invariants in the transverse transport
properties of cold atomic systems. It appears that fermionic atoms submitted to artificial
Abelian or non-Abelian gauge potentials and perturbed by an external forcing should
exhibit an integer quantum Hall-like effect. We have here shown that the
quantization of the transverse conductivity is robust if the gauge potential is set non-Abelian. The different values of
the Hall-like conductivity are computed for a specific $U(2)$ system
which depict a fractal phase diagram that should partly resist
perturbations encountered in a realistic setup. Cold atom systems
trapped in optical lattices present the versatility propitious to the
study of the fractional quantum Hall effect (FQHE) 
\cite{Sorensen2005}. These systems also constitute good candidates
for the exploration of the quantization and fractal structures we
have here emphasized.  We can speculate that the quantization 
of the transverse conductivity of neutral fermionic cold atom currents 
could open a new window in metrology.

\acknowledgments
N. G. thanks the F.~R.~I.~A. for financial
support. This research is financially supported by the
''Communaut\'e fran\c
caise de Belgique'' (contract ''Actions de Recherche Concert\'ees''
No. 04/09-312) and the F.~N.~R.~S. Belgium (contract F. R. F. C. No.
2.4542.02).

\end{document}